\documentclass{sig-alternate-10pt}
\usepackage[margin=1in]{geometry}
\usepackage{tikz}

\usepackage{multicol}
\usepackage{color}
\usepackage{xspace}
\usepackage{enumerate}
\usepackage{color}
\usepackage{url}

\usepackage{multirow}
\usepackage{graphicx}
\usepackage{comment}

\newcommand{\system}{\texttt{TStreamLLM}\xspace}

\usepackage{graphicx}
\graphicspath{{./figure/}}
\usepackage{subfig}

\usepackage{hyphenat}
\usepackage[utf8]{inputenc}
\usepackage{xcolor}
\usepackage[hidelinks]{hyperref}
\hypersetup{%
  colorlinks=false,
}
\usepackage{xspace}
\usepackage{cleveref}
\crefname{section}{§}{§§}

\makeatletter
\newcommand\subparagraph{%
  \@startsection{subparagraph}{5}
  {\parindent}
  {3.25ex \@plus 1ex \@minus .2ex}
  {-1em}
  {\normalfont\normalsize\bfseries}}
\makeatother
\usepackage{titlesec}
\titlespacing{\section}{4pt}{3pt}{3pt}

\newcommand{\xianzhi}[1]{{\textcolor{purple}{#1}}}	

\newcommand{\Ca}{\hyperref[Label:C1]{\textsterling$A$}\xspace}
\newcommand{\Cb}{\hyperref[Label:C2]{\textsterling$B$}\xspace}
\newcommand{\Cc}{\hyperref[Label:C3]{\textsterling$C$}\xspace}

\begin{document}
\pagestyle{empty}

\title{Harnessing Scalable Transactional Stream Processing for Managing Large Language Models [Vision]}
 
\author{
{
Shuhao Zhang,
Xianzhi Zeng,
Yuhao Wu, 
Zhonghao Yang
}\\
Singapore University of Technology and Design
} 
 
\maketitle

\begin{abstract} 
Large Language Models (LLMs) have demonstrated extraordinary performance across a broad array of applications, from traditional language processing tasks to interpreting structured sequences like time-series data. Yet, their effectiveness in fast-paced, online decision-making environments requiring swift, accurate, and concurrent responses poses a significant challenge. This paper introduces \system, a revolutionary framework integrating Transactional Stream Processing (TSP) with LLM management to achieve remarkable scalability and low latency. By harnessing the scalability, consistency, and fault tolerance inherent in TSP, \system aims to manage \textit{continuous \& concurrent} LLM updates and usages efficiently. We showcase its potential through practical use cases like real-time patient monitoring and intelligent traffic management. The exploration of synergies between TSP and LLM management can stimulate groundbreaking developments in AI and database research. This paper provides a comprehensive overview of challenges and opportunities in this emerging field, setting forth a roadmap for future exploration and development.
\end{abstract}

\section{Introduction}
\label{sec:introduction}
Large language models (LLMs) have become increasingly influential, propelling numerous advancements not just in natural language understanding and generation, but also in areas such as time-series analysis, structured sequence interpretation, and artificial intelligence overall~\cite{brown2020language, devlin2019bert, raffel2020exploring}. Their unprecedented scale and complexity allow them to excel at zero-shot and few-shot learning tasks~\cite{brown2020language,mann2020language}, opening up diverse applications across a multitude of domains. However, the promising capabilities of LLMs come with their own set of challenges.

\paragraph{Continuous Model Updates (\Ca):}\label{Label:C1} The success of LLMs hinges on significant resource consumption and a heavy reliance on the pre-training process~\cite{strubell2019energy,henderson2020towards}. As a result, there exists a knowledge cutoff for LLMs. While the world continually evolves with new concepts, events, and trends~\cite{openai2023gpt4,luu-etal-2022-time}, LLMs stay static after their pre-training. Therefore, keeping them updated and maintaining their relevance and accuracy pose significant challenges~\cite{jang2022towards}.

\paragraph{Concurrent Model Updates and Usage (\Cb):}\label{Label:C2} The demand for real-world applications that require \textit{reliable and prompt} responses amidst intensive \textit{concurrent model updates and usage} presents another layer of complexity. Addressing the requirement for concurrent model updates and usage is not only critical but also inevitable, as potential conflicts and dependencies among multiple services may arise.

\paragraph{Optimization and Acceleration (\Cc):}\label{Label:C3} Various techniques have been developed to accelerating model train and inference, such as mixed precision training~\cite{micikevicius2018mixed}, distillation~\cite{hinton2015distilling}, pruning~\cite{han2015learning}, and quantization~\cite{hubara2017quantized}. Additionally, the exploitation of novel hardware architectures~\cite{davies2018loihi} can enhance the performance of LLMs without significantly sacrificing their accuracy. However, adapting these methods for real-time operation and ensuring their compatibility with other concurrent services presents a significant challenge.

To address these issues, we introduce a visionary approach in this paper: \system. This innovative framework aims to achieve ultra-scalability and low latency in managing concurrent LLM updates and usage. The key concept behind \system is the integration of transactional stream processing (TSP) techniques~\cite{mao2023morphstream} into LLM management. TSP, an emerging data stream processing paradigm, offers real-time adaptation, data consistency, fault tolerance, and fine-grained access control~\cite{hirzel2014catalog}—qualities that make it suitable for managing LLMs under intensive concurrent stream processing scenarios~\cite{tatbul2017streaming}.

By leveraging TSP's scalability, fault-tolerance, and streaming semantics, \system empowers LLM management to substantially improve upon existing solutions. For instance, it reduces the best achievable long-run latency to a linear function of the single-user-single-run model manipulation overhead. 
These innovations could expand the potential of LLMs across a multitude of AI applications. Furthermore, the TSP-empowered LLM management system presents the database research community with flexible, adaptive methods for data ingestion, manipulation, and mining.

\textbf{In summary}, this paper makes the following contributions: We start by illustrating two practical use cases of LLMs, highlighting the pressing need for a system that can effectively manage continuous model updates, handle concurrent model updates and usage, and optimize and accelerate model operation in a real-time, scalable, and efficient manner (Section~\ref{sec:usage}). Next, we introduce our novel solution to these challenges: the \system framework. \system integrates TSP techniques into LLM management, offering potential improvements in efficiency, scalability, and adaptability (Section~\ref{sec:architecture}). Lastly, we explore the challenges and open research questions in this emerging field (Section~\ref{sec:challenges}). Our discussion sets a foundation for future research aimed at developing novel LLM architectures and management strategies leveraging TSP, thereby propelling advancements in AI and database research (Section~\ref{sec:conclusion}).

\section{Use Cases}
\label{sec:usage}
In this section, we delve into two significant real-world applications of \system, namely \textit{Real-time Patient Monitoring in Healthcare} and \textit{Traffic Management in Smart Nation}, showcasing how \system effectively tackles the three main challenges of LLM management (\Ca, \Cb, and \Cc).

\begin{figure}[t]
\centering
\includegraphics*[width=0.48\textwidth]{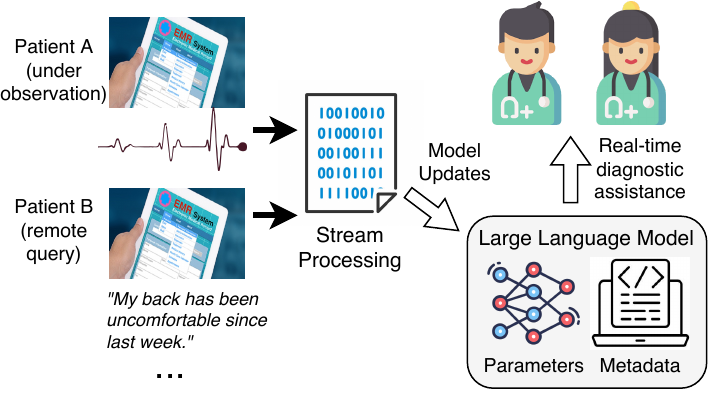}
\caption{\system applied in real-time patient monitoring in healthcare.}
\label{fig:llm_case1}
\end{figure}

\paragraph{Use Case 1: Real-time Patient Monitoring in Healthcare:}
Real-time patient monitoring has gained substantial relevance in the rapidly evolving field of healthcare~\cite{kakria2015real, kang2018recent}. A patient monitoring system implemented on \system enables the processing of a wide range of data, including electrocardiogram reports for patients under observation and medical condition descriptions from remote patients. By learning and analyzing these input data using the LLM, \system generates real-time health monitoring outputs and offers diagnostic assistance to doctors, as depicted in Figure~\ref{fig:llm_case1}.

To stay updated on the latest health condition of patients (\Ca), \system continuously fine-tunes the LLM to incorporate the most recent health data. By leveraging stream processing, the system efficiently carries out noise removal, feature extraction, and identification of key health indicators on input data. It concurrently updates LLM states (model parameters and metadata) using parallel executors, effectively meeting the real-time operational requirements (\Cc).

However, ensuring consistency in the LLM during concurrent model updates and queries poses a notable challenge (\Cb) due to the intricate dependencies involved in model access requests. \system successfully addresses this challenge by employing transactional concurrency control mechanisms. This allows for real-time querying and seamless access to the dynamically evolving LLM without impeding its ongoing training process, ensuring the efficient provision of diagnostic assistance to doctors.



\begin{figure}[t]
\centering
\includegraphics*[width=0.45\textwidth]{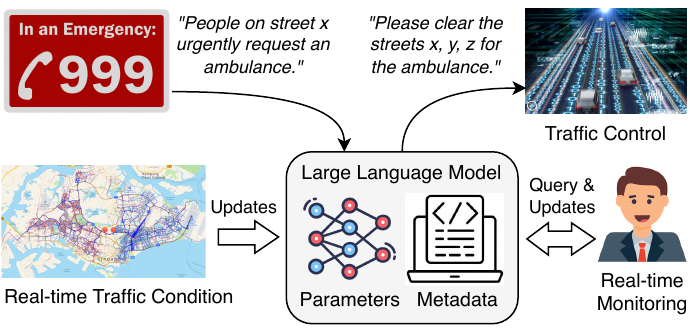}
\caption{\system's role in online traffic management within a smart nation framework.}
\label{fig:llm_case2}
\end{figure}

\paragraph{Use Case 2: Intelligent Traffic Management in Smart Cities:}
In the context of smart city traffic management, the optimization of city-wide traffic flow and response times necessitates an intelligent solution~\cite{putra2018intelligent, djahel2014communications}. However, there are challenges (\Ca and \Cb) posed by the dynamic nature of traffic data. These challenges involve maintaining model consistency and facilitating continuous learning in the face of data from diverse sources, such as road sensors, traffic cameras, and user-reported incidents.

\system excels in managing concurrent data streams, ensuring the LLM is consistently updated with real-time traffic conditions (Figure~\ref{fig:llm_case2}). Additionally, it collaborates with manual monitoring to handle complex traffic queries and offer context-aware recommendations (\Cc).

During emergency situations like ambulance requests, \system effectively demonstrates its real-time capabilities by promptly notifying the nearest ambulance, identifying the optimal route to the hospital, and simultaneously generating traffic control signals to facilitate the ambulance's movement. In more complex scenarios involving concurrent emergency calls (\Cb), \system efficiently learns and generates optimal traffic control strategies. It effectively allocates resources and prevents further damage.

\section{Harnessing Transactional Stream Processing for LLM Management}
\label{sec:architecture}
\begin{figure}[t]
\centering
\includegraphics[width=0.48\textwidth]{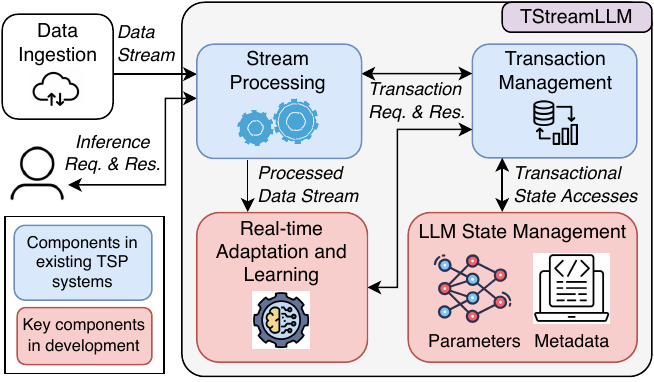}
\caption{Architectural overview of \system.}
\label{fig:llm_tsp_architecture}
\end{figure}

This section provides an overview of how \system harnesses the power of TSP to manage LLMs effectively. As illustrated in Figure~\ref{fig:llm_tsp_architecture}, \system uniquely integrates TSP techniques into LLM management, marking a pioneering framework that opens up avenues for future research.

\system is designed around four critical components: (1) \textit{Stream Processing} that efficiently processes real-time data streams and user inference requests, (2) \textit{Real-time Adaptation and Learning} that facilitates dynamic adaptation of the LLM based on incoming data, (3) \textit{Transaction Management} that guarantees model consistency and efficient update propagation, and (4) \textit{LLM State Management} that ensures the LLM remains up-to-date, managing the storage of LLM parameters and metadata. These components not only interlink to form the integrated \system, but also function independently, offering \system remarkable versatility across various scenarios.

\subsection{Stream Processing} \label{Stream_Processing}
The Stream Processing component is at the core of \system, designed to efficiently handle and process real-time data streams, supporting the optimization and acceleration of LLM under concurrent services (\Cc). As a plethora of data from user interactions, device logs, or sensor readings continuously flow in, this component acts as a dynamic dispatcher. It preprocesses the data, filters out irrelevant content, transforms raw data into a format digestible for LLMs, and performs various aggregations to distill meaningful insights.

The Stream Processing component utilizes advanced techniques to effectively manage high-velocity and high-volume data streams. To optimize the handling of incoming data, data stream compression~\cite{XianzhiReport22} is implemented, reducing storage and computational demands. Additionally, parallel processing~\cite{BriskStream} enables simultaneous management of multiple data streams, enabling \system to keep up with the constant influx of data.

The Stream Processing component goes beyond its role of handling incoming data streams for model updates and also addresses real-time user inference requests using transactional semantics. It efficiently processes and models user requests as transactions and delivers real-time responses based on the adapting LLM, facilitating seamless interaction between users and the Transaction Management component (Section~\ref{Transaction_Management}).



\subsection{Real-time Adaptation and Learning}
\label{Real_Time_Adaptation_and_Learning}
The Real-time Adaptation and Learning component plays a crucial role in the continuous fine-tuning of LLM (\Ca). It integrates with the Transaction Management component (Section~\ref{Transaction_Management}) to consistently retrieve the latest version of LLM parameters and metadata, and refine these states based on the insights derived from the processed data streams. This continuous learning mechanism allows the LLM to persistently enhance its performance and accuracy, maintaining relevance in the ever-evolving data landscape.


To efficiently perform real-time adaptation and improvement on the model, the Real-time Adaptation and Learning component utilizes concepts from Online Learning (OL), which is a machine learning technique that allows models to be incrementally updated as new data arrives without waiting for large batches~\cite{aljundi2019gradient}. OL enables the LLM to adapt to real-time changes in the continuous data stream, making it highly responsive to the shift in data stream patterns with minimal usage of computation resources, and supports rapid deployment in real-time decision-making scenarios.

However, ensuring the consistency of LLM states in the presence of concurrent inferences and model updates presents a significant challenge (\Cb). To address this, the Real-time Adaptation and Learning component models upstream state access operations as transactions, each transaction encapsulates a series of model update operations that must be performed jointly as an atomic unit. These transactions are subsequently handed over to the Transaction Management component (Section~\ref{Transaction_Management}) for reliable execution.

\subsection{Transaction Management} \label{Transaction_Management}


The Transaction Management component of \system plays a crucial role in ensuring data consistency and enabling efficient update propagation within a transactional stream processing framework~\cite{mao2023morphstream}. It is responsible for guaranteeing the correctness of LLM states in the presence of concurrent model updates and usages (\Cb). By incorporating transactional semantics into LLM state access management, \system ensures isolation among concurrent transactions, enabling their execution without interference. Furthermore, it ensures the durability of state updates, making them permanent even in the face of unexpected system failures.

To manage the execution of transactional requests received from upstream components (Section~\ref{Stream_Processing} and~\ref{Real_Time_Adaptation_and_Learning}), the Transaction Management component employs various concurrency control techniques, aiming to allow multiple transactions to proceed without locking any shared states. It carefully analyzes and resolves dependencies among state access operations within transactions. Subsequently, it adaptively schedules state access workloads to parallel executors, which then interact with the LLM State Management component for execution.

\subsection{LLM State Management} \label{LLM_State_Management}



The LLM State Management component manages the storage of shared stateful objects in LLMs, including parameters (e.g., word embeddings, weights, and biases) and metadata (e.g., training history, model hyperparameters). These states are continuously updated through transactions propagated from the Transaction Management component, ensuring that the LLM remains aligned with the latest insights derived from incoming data streams.

Scalability and efficiency are prioritized by the LLM State Management component, which is crucial for handling large language models that can comprise billions of parameters. To achieve this, \system employs a distributed storage strategy, where the LLM states are partitioned and distributed across multiple nodes. This approach harnesses the power of parallel computing, enabling the system to effectively manage and update LLM states while enhancing scalability.

Additionally, the LLM State Management component incorporates efficient indexing strategies to facilitate rapid retrieval and updates of model states. Techniques such as hashing and trie-based index structures are employed to expedite access to state objects, particularly in highly concurrent environments. These indexing techniques contribute to improved performance and efficient handling of LLM states within the system.

\section{Open Challenges and Opportunities}
\label{sec:challenges}
While \system demonstrates promising potential, there are still challenges and opportunities for research. 

\subsection{Scalable Stream Processing}
To effectively handle high-velocity data streams and update LLMs with minimal latency under high levels of parallelism and heavy workloads, it is crucial to enhance the scalability of \system. This challenge opens several avenues for future research:

\paragraph{Data Partitioning and Load Balancing:}
Effective data partitioning strategies can evenly distribute language model training data across parallel processing units, resulting in efficient resource utilization and minimized processing bottlenecks~\cite{rajbhandari2020zero}. Moreover, designing custom accelerators, GPUs, and multicore processors optimized for parallel processing and stream management can substantially enhance the scalability of stream processing. Future research should also investigate dynamic load balancing mechanisms that can adapt resource allocation in real-time according to fluctuating data rates and computational demands of the language models.

\paragraph{Domain-Specific Stream Processing:}
Integrating domain-specific knowledge~\cite{gu2021domain,gururangan-etal-2022-demix} into the stream processing pipeline enhances the efficiency of LLM management. Research can target developing bespoke stream processing operators and algorithms tailored to particular applications. Machine learning approaches could inform adaptive query optimization techniques that adjust execution plans based on incoming data stream characteristics, language model requirements, and resource availability. A critical challenge lies in managing the volume of training data processed, stored, and transmitted. Implementing domain-specific data stream compression techniques~\cite{XianzhiReport22} and approximation algorithms could economize resource consumption by accepting a trade-off between accuracy and reduced processing time.

\paragraph{Fault Tolerance and System Reliability:}
Maintaining system robustness is vital for \system, given the complexity and high volume of data processed by LLMs. Efficient recovery techniques like checkpointing, logging, and rollback mechanisms are essential to minimize disruptions, ensure system availability, and handle transaction failures. Approximate Fault Tolerance (AFT)~\cite{AF-Stream,10.1145/2505515.2505753} offers a promising approach, balancing error bounds and backup overheads while bolstering system stability. Future research should explore the potential of emerging hardware technologies and domain-specific fault tolerance strategies to improve system performance and ensure the scalability and reliability of \system in managing concurrent LLM updates.

\subsection{Real-time Adaptation and Learning}
Ensuring the relevance and accuracy of LLMs in the face of dynamic data streams and rapidly changing application requirements necessitates real-time adaptation and learning capabilities in \system. Future research can address this challenge by focusing on the following aspects:

\paragraph{Stream Data Selection:}
In an environment with large-scale, high-velocity data streams, LLMs face the challenge of selecting the most pertinent data for training~\cite{zhou2023stream,10.1145/3514221.3517836}. Traditional learning scenarios provide pre-defined datasets for incremental or transfer learning~\cite{jin2021lifelong,qin2022elle}, but this approach becomes unfeasible in a dynamic data stream environment~\cite{margatina2023dynamic}. Instead, the model needs to make knowledgeable decisions about data selection based on its existing knowledge. This challenge becomes evident during a newsworthy event, where the model is inundated with redundant information from various media outlets and online comments. In such cases, the model must adopt appropriate data selection techniques to balance the training volume with its understanding of the event, all the while maintaining its neutrality and objectivity.

\paragraph{Continual and Transfer Learning:}
In the realm of continual learning, catastrophic forgetting presents a significant challenge~\cite{wang2023comprehensive,gama2013evaluating}, whereas, in transfer or stream learning, the model's adaptability is of greater concern. In \system, both these issues co-exist, implying a need for models to possess both forward and backward transfer capabilities. This duality presents new challenges for existing methods. Given the continuous data stream, storing new knowledge becomes difficult for the model. When the model undergoes a training cycle, it can struggle to retain current factual knowledge, a problem exacerbated by random masking mechanisms employed during the training of models like BERT~\cite{guu2020realm}. In an online setting, with continuous data streams and no distinct task or domain boundaries, most offline continual learning methods fall short. Moreover, implementing a core set for replay through methods like gradient computation is particularly challenging for LLMs, leading to potentially high costs.

\paragraph{Adaptive and Efficient Model Training:}
The traditional static nature of LLMs post-pre-training presents challenges in dynamic real-world scenarios. The \system framework emphasizes the need for frequent, accurate model updates with reduced latency. Classic model updating, involving steps like forward propagation, loss computation, back propagation, and parameter updates, can introduce system latency due to its sequential nature. To address this, we suggest predictive and concurrent model training methods. These would include predicting upcoming loss values based on previous ones, enabling continuous updates even before the completion of prior ones. Another promising direction involves preemptive identification and updating of parameters requiring significant changes post-loss computation, aiming to avoid potential conflicts.

\subsection{Streaming Transaction Processing}
The \system framework hinges on the effective transactional management of LLMs in response to high-velocity data streams. The dynamic nature of data streams and the necessity for real-time response pose exciting research challenges in the realm of streaming transaction processing:

\paragraph{Transactional Model Updates:}
Incorporating concurrent and predictive model training methodologies introduces several challenges, including maintaining ACID properties during concurrent updates, especially with high-velocity data streams. Concurrent updates can also create potential conflicts and dependencies among services, adding complexity. Therefore, future research should develop efficient conflict detection and resolution strategies specific to LLMs. Despite the challenges, these strategies, transactional guarantees, and conflict resolution mechanisms could significantly enhance model training efficiency and concurrent update management in the \system framework, improving its effectiveness and reliability.

\paragraph{Scalability and Performance Trade-offs:}
As the demand for LLMs in real-time applications grows, the \system framework must be capable of processing transactions efficiently under high loads. Future research could investigate strategies for scaling streaming transaction processing capabilities~\cite{mao2023morphstream} to accommodate growing volumes of data streams. This could involve exploring innovative parallel processing techniques, distributed computing solutions, or the use of emerging hardware technologies to accelerate transaction processing. Furthermore, there may be trade-offs between transaction processing speed, system consistency, and model accuracy. Understanding these trade-offs, and developing strategies to balance these conflicting demands could be another crucial area of exploration.

\subsection{LLM State Management}
The ability to manage the state of LLMs effectively within the \system framework forms a critical component of maintaining updated and consistent LLMs. This state management plays a pivotal role in ensuring the framework's real-time response capabilities. The areas of investigation worth delving into within this domain include:

\paragraph{State Storage and Version Control:}
Storage efficiency in LLM state management demands the exploration of innovative methods for compression and storage optimization. Techniques such as delta encoding~\cite{douglis2003application} and sparse representations~\cite{zhang2015survey} could minimize storage requirements, thus enhancing the scalability of the \system framework. Moreover, contemplating the future integration of vector data management systems~\cite{wang2021milvus, guo2022manu} into the LLM State Management could further optimize storage and retrieval operations, despite the inherent challenges in handling high-dimensional data. In addition, managing different versions of LLM states efficiently in \system, while minimizing the overhead of maintaining these versions, is of paramount importance. This calls for efficient versioning and snapshotting techniques enabling access to and querying of previous LLM states, which in turn contributes to the robustness and reliability of the system in various use cases.

\paragraph{Optimization and Security Assurance:}
LLM state management significantly impacts system performance and resource utilization. 
Optimizing elements such as memory hierarchies, storage systems, and processing resources can significantly enhance LLM performance and the overall scalability of the \system. The balance between resource utilization and system performance should remain a priority. Security and privacy constitute another critical facet of LLM state management, as they prevent unauthorized access and protect the model from potential damage. Future research should concentrate on devising privacy-preserving techniques for data processing and LLM adaptation, such as federated learning~\cite{li2021survey} and differential privacy~\cite{10.14778/3476249.3476277}. These methods could protect sensitive data while enabling LLMs to learn from various data sources, thereby ensuring the \system framework complies with privacy standards without compromising its learning capabilities.

\section{Conclusion}
\label{sec:conclusion}
In this paper, we introduce a novel perspective on merging transactional stream processing with LLMs management, setting forth a promising research trajectory. We outline key challenges and potential solutions in areas such as scalable stream processing, real-time adaptation and learning, streaming transaction processing, and LLM state management. This integration aims to solve challenges related to data selection, continual learning, and efficient model training in a high-velocity data stream environment. We propose new strategies for transactional model updates, emphasizing concurrent and predictive model training to mitigate system latency and conflict resolution issues. We emphasize the necessity to respect ACID properties and tackle potential service conflicts in high-load applications. We also spotlight the importance of fault tolerance and system reliability for the \system framework to handle high-volume data processed by LLMs effectively. Our vision presents the possibility of revolutionizing the management of LLMs. By addressing the open challenges we've outlined, we hope to inspire further innovation, leading to the development of robust, efficient, and scalable solutions in this rapidly evolving field.

\bibliographystyle{abbrv}
\bibliography{mybib}

\end{document}